\documentclass[twocolumn,nofootinbib,showpacs,preprintnumbers,amsmath,amssymb]{revtex4}
\usepackage{graphicx}
\newcommand{\be}{\begin{equation}} \newcommand{\ee}{\end{equation}} 
\newcommand{\ba}{\begin{eqnarray}} \newcommand{\ea}{\end{eqnarray}}

\newcommand{\bea}{\begin{eqnarray}}
\newcommand{\eea}{\end{eqnarray}}
\begin{document}
\preprint{UCB-PTH-09/25}
\preprint{MCTP-09-44}
\newcommand{\we}{\wedge}
\title{A Non-thermal WIMP `Miracle'}
\author{Bobby Samir Acharya\footnote{bacharyaATcern.ch}}
\affiliation{Abdus Salam International Centre for Theoretical
Physics, Strada Costiera 11, Trieste, Italy\\and\\INFN, Sezione di Trieste}
\author{Gordon Kane\footnote{gkane@umich.edu} and Scott Watson\footnote{watsongs@umich.edu}\footnote{On leave from Department of Physics, Syracuse University, Syracuse, NY 13244}}
\affiliation{
Michigan Center for Theoretical Physics, Ann Arbor, MI}
\author{Piyush Kumar\footnote{kpiyush@berkeley.edu}}
\affiliation{Berkeley Center for Theoretical Physics
University of California, Berkeley, CA 94720\\and\\
Theoretical Physics Group
Lawrence Berkeley National Laboratory, Berkeley, CA 94720}
\vspace{0.5cm}

\date{\today}

\vspace{0.3cm}

\begin{abstract}
Light scalar fields with only gravitational strength couplings are typically present in UV complete theories of physics beyond the Standard Model.
In the early universe it is natural for these fields to dominate the energy density, and their subsequent decay -- if prior to BBN -- will typically yield some dark matter particles in their decay products.  In this paper we make the observation that a Non-thermal WIMP `Miracle' may result:  
that is, in the simplest solution to the cosmological moduli problem, non-thermally produced WIMPs can naturally account for the
observed dark matter relic density. Such a solution may be generic in string theory compactifications.
\end{abstract}
\maketitle
\newpage
\vspace{-1.2cm} 

\section{Introduction}

For several decades, a compelling theoretical picture for dark matter 
has been developed and widely applied,
based on what is now called the Thermal WIMP `Miracle'. In this
picture, the early Universe is very hot and dense and essentially all particle species are
in thermal and chemical equilibrium. As the Universe expands and cools to
a temperature of order the mass of the dark matter particle $\chi$, the annihilations
of $\chi$ cease to be efficient at reducing the number of particles compared to the cosmic expansion and a `freeze-out' occurs. The
resulting relic number density of $\chi$-particles then depends only on the ratio of the annihilation cross-section of $\chi$
and the Hubble scale near the freeze-out temperature. For Weakly Interacting Massive Particles (WIMPs), i.e.
particles with electroweak interactions and masses,
the `miracle' is that this ratio is in good agreement with that deduced from astrophysical and cosmological experiments (see \cite{DMReview} for reviews).

Whilst this is an extremely compelling idea, it can often be difficult to implement
in practice, with the relic density predicted by theory disagreeing with the data by
a couple orders of magnitude in both directions. That is, the thermal WIMP `miracle' 
faces significant challenges when directly confronted with precision data. It is not clear at present if this should be viewed
as a failure of the theoretical models or as a phenomenological guide to select particular classes of models. 
For example, in supersymmetric models, where the thermal relic idea has perhaps been most extensively explored,
most of the parameter space yields an incorrect dark matter density, prompting attempts to look at 
special regions \cite{special}.

Furthermore, by considering the high energy behavior (UV completion) of phenomenologically based models, 
it becomes less straightforward to motivate cosmologies in which the Universe 
is in thermal equilibrium prior to Big Bang Nucleosynthesis (BBN). This
is partly because many UV completions lead to the inclusion of moduli -- neutral scalar 
fields which couple to matter only gravitational -- 
and for a wide range of masses these moduli will typically evolve to dominate the cosmic energy density prior to BBN.

As a result, the decay of the moduli must give rise to temperatures 
above that of Big Bang Nucleosynthesis (BBN) - a few MeV -  or the predictions of BBN will be ruined. 
This is the cosmological moduli problem (formerly known as the Polonyi problem) \cite{Coughlan:1983ci,Ellis:1986zt,de Carlos:1993jw,Banks:1993en}
and typically requires that the moduli masses are at least
10 TeV or greater.  Of course, exceptions are possible. A period of low-scale inflation could dilute the moduli density \cite{Lyth:1995ka}.  Another possibility is if signigicant energy is not stored in the fields.  This could be accomplished by dynamics, or if the moduli are stabilized initially near points of enhanced symmetry \cite{esp,dineus}.  However, in this case, the need for a perturbative low energy theory necessarily requires at least one modulus to remain unprotected, and thus a substantial contribution to the energy density should still be expected \cite{dineus}. We will not discuss these interesting possibilities further here.

Moduli which evade the cosmological moduli problem have masses which are typically greater than that of
WIMP dark matter candidates,  so their decays will necessarily result in the production of some dark matter.  
In fact, the yield of dark matter particles can be quite large, 
since moduli tend to couple to the matter sector universally
with gravitational strength couplings ($\sim 1/{m_p}$) and the branching ratios to stable particles are expected
to be substantial. Thus, the density of dark matter at the time of moduli decay will typically be a sizable
fraction of the energy density of the moduli themselves. 
Non-thermally produced dark matter has been considered previously by many authors \cite{nonthermal,MoroiRandall,G2DM,Nakamura:2006uc}.

In this paper we will attempt to elaborate the connection between the moduli problem and 
non-thermally produced dark matter. 
In the absence of any strong guidance as to how one might proceed from
fundamental theory,
we will take the viewpoint of a {\it generic supergravity theory containing moduli}.
That is, we will simply take an effective field theory approach to the general problem
of moduli coupled to matter in a supersymmetric framework.
{\it A priori} though, in order to proceed, one still requires a clue about
the mass spectrum of superpartners, moduli and the dark matter candidate.  In fact,
the effective theory itself provides such a clue.

To see this, note that in a generic supergravity theory the moduli masses are
of order $m_{3/2}$ (the gravitino mass) and, furthermore, 
all scalars receive mass term contributions of the same order \cite{reviews}. 
This includes all squarks and sleptons. Thus, since $m_{3/2}$ is required to be at least 10 TeV
because of the moduli problem, squarks and sleptons are at least as massive. Further, as we will argue
below, gauginos need not be as massive as the squarks and sleptons in such a generic supergravity theory. On the other hand,
in the absence of special symmetries, higgsinos also generically have a mass of 
order $m_{3/2}$ by the Giudice-Masiero mechanism. If we thus consider the 
simplest model within this framework - the minimal supersymmetric standard model (MSSM) coupled to moduli, these
arguments suggest that $\chi$ is a (neutral) gaugino, i.e. a Bino, Wino or
a mixture thereof.

Rather surprisingly, within the class of models with spectra suggested by the above arguments,
there exists a Non-thermal WIMP `Miracle' under very general conditions. In other
words, by solving the cosmological moduli problem, one automatically obtains a consistent solution to the
basic dark matter problem with WIMPs when the moduli masses are 10's of TeV..
However, as will be seen, compared to the thermal case, the WIMP
annihilation cross-section is larger, thereby suggesting regions of parameter space previously
avoided based on the thermal dark matter picture. 

Examples of models with the sorts of spectra considered here have been considered previously
in the context of AMSB \cite{MoroiRandall} and more recently in the context of
$M$ theory \cite{G2MSSM, G2DM} where the non-thermal production of dark matter was
emphasized (see also \cite{heckman}). These examples show that with additional, well motivated assumptions, that the
arguments of this paper can be sharpened and detailed models can be constructed.
We emphasize that many of the ideas given here were inspired by the seminal work of \cite{MoroiRandall}.

\section{Moduli Masses and Supersymmetry breaking}
The moduli fields must enter the supergravity potential. If not, they would remain
massless and not gain the vevs required to explain, e.g. the value of the fine-structure constant $\alpha_{em}$ 
or the Yukawa couplings in the Standard Model.  If $V$ is the potential for all scalar fields, and the only supersymmetry
breaking physics is generated at the scale of order $\sqrt{F}$, then the moduli masses
will generically also be of order the gravitino mass as we now review.

The supergravity potential evaluated in the vacuum is of the form
\be \label{eqn1}
V = e^{K/m_p^2} F^i F_i -3m_{3/2}^2 m_{p}^2
\ee
where $F_i$ are the vacuum values of the SUSY-breaking $F$-terms which are to be summed over all scalars $\phi_i$ with non-trivial $F$-terms, $K$ is the Kahler potential, and $m_{3/2}$ is the gravitino mass.  
Hence, because the observed vacuum energy today
is so small,
\be
m_{3/2} \sim F/m_{p}
\ee
where $F$ is of order the dominant $F$-term. This must be true, regardless of the mechanism
of supersymmetry breaking. In fact, $F/m_{p}$ sets the typical mass scale for all
scalar fields appearing in $V$ in a generic supergravity theory. For instance,
since
\be \label{eqn2}
F_i = {\partial W \over \partial{\phi_i}} + {\partial K \over \partial{\phi_i}} W \equiv \partial_{i} W + K_i W
\ee
where $W$ is the superpotential, there are terms in $V$ of order
\be
V \sim K_i K^i {|W|^2 \over m_p^2}. 
\ee
If we consider the terms in $K$ (or in $K_i K^i$ ) of order $\phi_i \phi_i^*$ 
we obtain contributions to $V$ like
\be
V \sim \phi_i \phi_i^* {|W|^2 \over m_p^2}
\ee
In the vacuum $|W|^2 \sim m_{3/2}^2 m_{p}^4$ hence these terms are mass terms for $\phi_i$ of
order $m_{3/2}$. Hence, all scalars receive contributions to their masses
of order $m_{3/2}$ in a generic supergravity theory. This includes moduli as well as matter
scalars such as higgses, squarks and sleptons \footnote{See \cite{BSA} for examples.}.

Exceptions exist to the above general statements. In the language of low energy supersymmetry, these would correspond
to very special vacuum Kahler and superpotentials for the moduli and matter fields. For instance,
if the moduli dynamics is $R$-symmetry preserving, then the moduli potential will not break supersymmetry
and the moduli could, in principle, obtain large masses and vevs independent of the value of the
gravitino mass, whose value is set by additional $R$-breaking dynamics at another scale.
We will not consider such exceptions further, except in the conclusions, 
and will adopt the viewpoint of the `generic supergravity Lagrangian'.

Notice also that, if there were a bare, large mass term in $W$ for the moduli, the fact that the
moduli get vevs will give $W$ a vev in general and therefore contributes to $m_{3/2}$ thereby
connecting again the moduli and gravitino mass. Again, to avoid this requires $R$-symmetric
moduli dynamics.

\subsection{Gauge Mediation}
In gauge mediation the dominant
$F$-term $\sqrt{F} \leq 10^{10} $ GeV, though typically $F$ is taken to be
much smaller than what the upper limit (of high scale gauge mediation) suggests.
Hence, the gravitino mass and moduli masses are of order
\be
m_{3/2} \sim m_{\phi} \sim F/m_{p} \leq 10 \mathrm{GeV}
\ee

Since theories of gauge mediated supersymmetry breaking have $\sqrt{F}$ between a TeV and
$10^{10}$ GeV they
lead to moduli masses in the wide range:
\be
10^{-3} \mathrm{eV} \leq m_{\phi} \leq 10\; \mathrm{GeV}
\ee

Of course, in gauge mediation, charged scalars get much larger corrections to their masses
from their interactions with the `messengers of SUSY breaking', but the moduli do
not.

This range of moduli masses leads to cosmological problems. First of all, since the
moduli have Planck scale suppressed couplings to matter, 
this range of masses gives moduli whose lifetimes are between 100 and $10^{42}$ years.
After inflation, when the Hubble scale becomes of order $m_{\phi}$
the moduli begin to oscillate and will quickly dominate the energy density of the universe
over radiation. This range of masses corresponds to temperatures 
\be
\mathrm{TeV} \leq T_{osc} \leq 10^{10} \; \mathrm{GeV}
\ee
which means that such moduli would begin to dominate {\it before} BBN. Therefore, such models
lead to Universes which are dominated by moduli for very long periods. Further, after the moduli decay
the Universe is not reheated enough to start BBN with the correct conditions. 
Hence, gauge mediation models coupled to moduli require quite special Kahler and super potentials which
would allow for the moduli to be very massive compared to the gravitino.

\subsection{Moduli Masses in Gravity Mediation}

In gravity mediation,  $\sqrt{F}$ is much higher,
e.g. $10^{11}$GeV to $10^{16}$GeV and 
hence $m_{3/2}$
is usually taken to be of order TeV, since supersymmetry is assumed to solve the
hierarchy problem.
Then, through the typical supergravity couplings
discussed above,
the moduli will also end up with TeV scale masses. Such moduli will decay {\it during}
BBN and will typically ruin its successful predictions for light element abundances.
The gravitino also leads to similar problems. This is just the usual moduli problem.

In gravity mediation, there is a simple solution to these
problems. One can simply raise the scale of SUSY breaking ($\sqrt{F}$) by a factor of a few which
raises both $m_{3/2}$ and $m_{\phi}$ by one order of magnitude. 
This decreases the moduli
lifetime by three orders of magnitude and is consistent with BBN occurring just after
the moduli have decayed. A detailed model in this case has been described in 
\cite{G2DM}. Notice that the gravitino mass has also been raised above the
TeV scale, since $F$ is increased.

Since, as we have discussed above, all scalars, including squarks, sleptons and Higgses, will have masses of order $m_{3/2}$ 
the fine tuning problem between $m_{3/2}$ and $m_{Z}$ is naively much worse in theories with moduli than in the usual
little Hierarchy problem.

A crucial point, however, is that 
the gauginos do not have to be as massive as the squarks and sleptons:
they can easily be lighter if {\it the field whose $F$-term dominates
SUSY breaking in the hidden sector is not the field whose vev generates the gauge couplings}
$\alpha$. In fact, in the generic supergravity point of view, there is no reason why the
gauge coupling function should be dominated by the field with the dominant $F$-term. 
This can presumably also be understood as the consequence of an approximate $R$-symmetry in this sector.  
Here we emphasize that it is quite generic. However, in general this approximate $R$-symmetry only suppresses the gaugino masses but not that of 
the higgsinos. One could also have special $R$-symmetries in which both gaugino and higgsino masses are suppressed, as in split supersymmetry \cite{split}. 
Hence, we expect that the gauginos are significantly lighter than the squarks and sleptons. In the absence of 
special $R$-symmetries, we expect the higgsinos to get masses through the Giudice-Masiero mechanism of order $m_{3/2}$.

What we learn is --  by considering the simplest solution to the cosmological moduli problem, we
obtain a rough picture of the spectrum of BSM particles.
Existing models which have this kind of spectrum include Anomaly Mediated Supersymmetry Breaking Models (AMSB) \cite{MoroiRandall} and the $G_2$-MSSM \cite{G2MSSM}.

\subsection{The Non-thermal WIMP `Miracle'}

By taking this effective field theory approach we have arrived at a picture which suggests 
that supersymmetry breaking is gravity mediated with moduli and gravitino masses of order 10 TeV.
The squark, slepton and higgsino masses are also of order 10 TeV, whereas the gauginos
are typically lighter. 
Cosmologically, these moduli inevitably dominate the Universe after inflation and up to BBN.

We can now ask: with the spectrum roughly fixed by these arguments, how much dark matter is produced, both thermally, 
and non-thermally by the decays of the moduli fields? In other words, is there a connection between the
moduli and dark matter problems?

After inflation, when the Hubble expansion becomes comparable to the mass of the moduli ($m_{\phi}$) they will 
begin to oscillate in their potential. The resulting energy density will dilute like normal pressure-less matter and they will quickly come to dominate the energy density.  The oscillations will begin at a temperature determined by the moduli mass
\be
T_{osc} \sim \left(m_{\phi} m_{p}\right)^{1/2}.
\ee
The resulting condensate will then decay when the expansion rate becomes comparable to the decay rate 
\be \label{modulidecay}
H \sim \Gamma_{\phi} \sim {m_{\phi}^3 \over m_{p}^2}.
\ee
Given that the moduli dominate before their decays, the resulting energy density is (assuming instantaneous decay)
$\rho_{decay} \sim {\Gamma_{\phi}}^2 m_{p}^2 = {m_{\phi}^6 / m_{p}^2}$ and so the 
expected number density of dark matter particles will be 
\be
n_{\chi} \sim Br_{\phi \rightarrow \chi } \left( \frac{\rho_{d}}{m_{\chi}} \right)
\sim Br_{\phi \rightarrow \chi } \left( { m_{\phi}^6 \over m_{\chi} m_p^2 } \right)
\ee
We can compare with the critical number density for annihilations to occur,
\be
n_c \sim {H \over \sigma v} \sim {\Gamma_{\phi} \over \sigma v} \sim \frac{m_\phi^3}{m_p^2 \langle \sigma v \rangle},
\ee
where $\sigma v$ is the self annihilation cross-section times velocity of the produced dark matter particles. 

\be
\frac{n_\chi }{n_c}  \sim Br_{\phi \rightarrow \chi } \left( \frac{m_\phi^3}{m_\chi} \right) \langle \sigma v \rangle 
\ee
Taking typical weak scale values for the dark matter mass ($\sim 100$ GeV) and cross-section ($\sim 10^{-24} \mathrm{cm}^3 \mathrm{s}^{-1} \sim 10^{-7}\mathrm{GeV^{-2}}$),
and moduli masses in the range to address the cosmological moduli problem ($\sim 10-100$ TeV), 
we find that the number density of produced particles is 
easily large enough for the $\chi$ particles to annihilate efficiently, unless the branching
ratio into dark matter particles is very small.  This is unlikely if dark matter particles 
are gauginos, because the gauge coupling is a modulus vev and this has an order one coupling to the
gauginos. They will thus continue to
annihilate until their number density becomes of order ${\Gamma_{\phi} \over \sigma v}$. This is
the non-thermal analogue of `thermal freeze out'. Hence, the non-thermal freeze out number density is
\be \label{num}
n_{\chi} \sim {H(T_{rh} ) \over \sigma v}
\ee
where the temperature of the Universe after the decay is
\be
T_{rh} \sim (\Gamma_{\phi}m_{p})^{1/2} \sim {m_{\phi}^{3/2} \over m_{p}^{1/2}}
\ee

The fact that a non-thermal freeze out actually occurs is by itself an important 
statement, making the relic density more model independent than if it did not
occur.
Assuming no further entropy production in the Universe after this stage, the
resulting relic abundance is 
\begin{footnotesize}
\be 
\Omega_\chi h^2 \approx 0 .1 \times \left( \frac{m_\chi}{100 \, \mathrm{GeV}} \right) \left( \frac{10.75}{g_\ast} \right)^{1/4} \left( \frac{\sigma_0}{\langle \sigma v \rangle} \right) \left( \frac{100 \, \mathrm{TeV}}{m_\phi}  \right)^{3/2}
\ee
\end{footnotesize}
where $\sigma_0 = 3 \times 10^{-24} {\mathrm{cm}^3}{\mathrm{s}^{-1}} $ and we have assumed\footnote{Larger values for the proportionality constant were found from explicit calculations in \cite{G2DM}, and smaller values are possible as well (see e.g. \cite{Nakamura:2006uc}).} the constant of proportionality in (\ref{modulidecay}) is around a factor of ten.
For $\chi$ particles with order $100$ GeV mass, a $100$ TeV moduli mass scale gives excellent
agreement with the data. Note that, (\ref{num}) expresses the number density at freeze-out as the ratio of the Hubble parameter to a particle physics cross-section, just as in the thermal case, except now the Hubble scale is evaluated at the moduli reheating temperature.  This is the non-thermal WIMP `miracle'. With larger moduli masses, smaller cross-sections are also possible. However, this separates
the Electroweak scale from the scale of supersymmetry breaking even further and, for
masses beyond 1000 TeV there are no good dark matter candidates in the spectrum whose cross-section
is small enough, leading to too little non-thermal dark matter.
Therefore, the preferred moduli mass scale is roughly 10 to 100 TeV.

For $\mathcal{O}(100)$ GeV masses, the $\chi$ particles freeze out at temperatures of order a few MeV, so the annihilation cross-section is about two to three orders of magnitude larger
than the thermal case, where freeze out occurs at a few GeV.
Also note that larger moduli masses dilute the final relic
density, so that if the dark matter is to be explained this way by gauginos, it requires moduli masses
just beyond the weak scale.

\section{Discussion}

The consistent solution of both the dark matter and moduli problems suggests that there
is another important scale in nature, of order 10-100 TeV. 
The picture which emerges here is,
in some sense, as compelling as the thermal case, but has the additional advantage that it is
consistent with string theory and other frameworks with moduli fields, without having to invoke
`unknown dynamics' which decouple the moduli.
The non-thermal dark matter picture requires larger annihilation
cross-sections than the thermal case, thereby suggesting a class of models which would not
be considered based upon the thermal relic density picture. 
Larger cross-sections are also helpful to explain the cosmic ray positron excess observed by PAMELA. 
For instance, the neutral Winos (with annihilation cross-section $\sigma v \approx 2.4 \times 10^{-24} \, {\mathrm{cm}^3}\cdot{\mathrm{s}^{-1}}$) which
are the dark matter particles in \cite{MoroiRandall, G2MSSM} could be a good fit to the
PAMELA data \cite{WinoPamela,Feldman:2009wv}. Non-thermally produced dark matter particles will in general
have a different phase space distribution than the thermal case; in principle this could lead to
observable consequences for structure formation and the cosmic microwave background. Finally, although entropy production by the decay of moduli also dilutes any pre-existing baryon asymmetry, the problem can be naturally solved if either a sufficiently large initial baryon asymmetry (such as by the Affleck-Dine mechanism) is generated \cite{Kumar:2008vs}, or if the decays of the moduli themselves generate both the baryon asymmetry and Dark Matter \cite{Kitano:2008tk}.

We also point out that these arguments tend to disfavor gauge mediated supersymmetry breaking.
If gauge mediated supersymmetry breaking were discovered at the LHC, it would imply that
our supergravity vacuum is extremely non-generic, in that it allows large moduli masses whilst keeping
the gravitino very light. An example might be a theory in which the moduli dynamics is $R$-symmetric
whilst supersymmetry breaking is not. This might allow a decoupling of $m_{3/2}$ and $m_{\phi}$.
However, as noted in \cite{Banks:1993en}, decoupling these scales is very difficult in general.

\acknowledgements
We would like to thank Paolo Creminelli, Guido D'Amico, Dan Grin, Jonathan Heckman, Shuntaro Nakamura, and Jorge Norena for useful discussions.
The work of P.K. is supported by the U.S. Department of Energy under
contract no. DE-AC02-05CH11231 and NSF grant PHY-04-57315.
The research of G.K. and S.W. is supported in part by the Department of Energy and the Michigan Center for Theoretical Physics.
S.W. would also like to thank U of T - Austin for financial support under National Science Foundation Grant No. PHY-0455649.

\end{document}